\documentclass{article} % For LaTeX2e
\usepackage{nips13submit_e,times}
\usepackage[utf8]{inputenc}
\usepackage[english]{babel}
 
\usepackage{natbib}
\usepackage{hyperref}
\usepackage{url}
\usepackage{graphicx}
\usepackage[]{amssymb}
\usepackage[]{amsmath}
\usepackage{float}
\usepackage{wrapfig}

\title{Rapid seismic domain transfer: \\ Seismic velocity inversion and modeling \\ using deep generative neural networks}
\author{
Lukas Mosser \\
Department of Earth Science and Engineering\\
Imperial College London\\
\texttt{lukas.mosser15@imperial.ac.uk} \\
\And
Wouter Kimman \\
Open Energi Ltd.\\
\texttt{wouter.kimman@gmail.com} \\
\And
Jesper Dramsch \\
Technical University of Denmark\\
\texttt{jesper@dramsch.net} \\
\And
Steve Purves \\
Euclidity SL\\
\texttt{steve@euclidity.com} \\
\And
Alfredo De la Fuente\\
Skolkovo Institute of Science and Technology\\
\texttt{alfredo.delafuente@skoltech.ru} \\
\And
Graham Ganssle\\
Expero Inc.\\
\texttt{graham.ganssle@experoinc.com} \\
}

\nipsfinalcopy % Uncomment for camera-ready version

\begin{document}

\maketitle

\begin{abstract}
Traditional physics-based approaches to infer sub-surface properties such as full-waveform inversion or reflectivity inversion are time consuming and computationally expensive. We present a deep-learning technique that eliminates the need for these computationally complex methods by posing the problem as one of domain transfer. Our solution is based on a deep convolutional generative adversarial network and dramatically reduces computation time. Training based on two different types of synthetic data produced a neural network that generates realistic velocity models when applied to a real data set. The system’s ability to generalize means it is robust against the inherent occurrence of velocity errors and artifacts in both training and test datasets.

\end{abstract}

\section{Introduction}

The task of inferring subsurface geological structures from depth-domain seismic data is a computationally demanding process that frequently appears in geophysical studies and hydrocarbon exploration. Typically, seismic inversion is performed by means of wave inversion methods of a simple prior model of the subsurface and using a backpropagation loop \citep{lailly1983seismic,tarantola1984inversion} to iteratively reduce the mismatch between the observed seismic data and the computed synthetic model \citep{pratt1998gauss, virieux2009overview}. Although this approach leads to satisfactory results in practice, it requires an overwhelming amount of computer resources with no guarantee of global convergence; making it inappropriate when time and computing constraints are strict or when we need to perform the same task for a number of geological scenarios.

As an alternative, we propose a data-driven approach that uses deep generative neural networks to formulate the seismic forward and inversion process as a domain transfer problem, which allows us to learn two functions from the datasets: \textsc{1)} a function to map from the seismic geo-model to the seismic amplitude domain \textsc{2)} a function to map from seismic amplitude to the geo-model domain. One of the main advantages of this approach comes from the fact that the training step of the algorithm does not require a set of paired input-output images in the dataset. 

We present examples of the resulting forward and inverted datasets using the domain transfer method, based on simple synthetic structural models, as well as the Marmousi 2D dataset. Finally, we highlight challenges and possible applications of the proposed approach.

\section{Theory}
Texture transfer or neural style transfer is an area of research in computer vision. \citet{gatys2016image} used an iterative process to transfer camera photographs into a desired artistic style. They showed results that extracted the features of pre-trained VGG networks \citep{2014arXiv1409.1556S} to model the desired output. This is a computationally expensive iterative process. However, \citet{johnson2016perceptual} specialise a single network per textural style, removing the need to solve an iterative minimisation problem. \citet{isola2016image} reframed the problem in the sense of a domain transfer problem. Here a generative model could be built that transfers the original data to the artistic style. Particularly, a generative adversarial network (GAN) was used with pair-wise corresponding images. \citet{zhu2017unpaired} loosened the constraint on pair-wise training data in a cycle-consistent GAN that learned transfer function between domains. Seismic inversion is an expensive iterative task similar to the computer vision problem discussed here. We use neural style transfer to find a transfer function from seismic amplitude data to velocity functions.  We show  that this process can benefit from the advancements in deep learning and computer vision.

Deep convolutional generative adversarial networks (DCGAN) consist of two powerful neural networks that learn by competition  \citep{goodfellow2014, Radford2016}. The generator network \textbf{G} draws samples from a noise prior or so-called latent space. The generated output is presented to the discriminator network \textbf{D} in a randomised switch with real data. The discriminator determines whether the output is generated by \textbf{G} or real. A loss function determines the rate at which both networks learn. In this case, \textbf{G} gets better at generating realistic outputs and \textbf{D} improves the ability to evaluate the realism of inputs. In a cycle-consistent setup, we train two GANs in parallel. The generator \textbf{G} learns the forward generative model. The second generator network \textbf{F} learns the inverse generative model. The GANs are set up to perform a full circle in the calculation. Input from domain \textbf{X} is mapped to domain \textbf{Y} by generator G, then generator \textbf{F} maps the result from domain \textbf{Y} to domain \textbf{X}. Ideally, the output of the cycle resembles the input so that $\textbf{F}(\textbf{G}(x_i)) \approx {x_i}$.

Both networks \textbf{G} and \textbf{F} are subject to an adversarial loss objective. The adversarial loss from the network of \citet{zhu2017unpaired} is defined as:
\begin{equation}
    \mathcal{L}(G,D_Y,X,Y) = \mathbb{E}_{y\sim p_{data}(y)} [\log D_Y(y)] + \mathbb{E}_{y\sim p_{data}(x)} [\log (1-D_Y(G(x)))] 
\end{equation}
with the second adversarial loss being equivalent as $\mathcal{L}(F,D_X,Y,X)$.

The cycle of the two GANs has to be consistent in the forward pass $x \rightarrow G(x) \rightarrow F(G(x)) \approx x$ as well as the backward pass $y \rightarrow F(y) \rightarrow G(F(y)) \approx y$. \citet{zhu2017unpaired} formalise the cycle consistency loss as follows:
\begin{equation}
    \mathcal{L}(G,F) = \mathbb{E}_{x\sim p_{data}(x)} [||F(G(x))-x||_1] + \mathbb{E}_{y\sim p_{data}(y)} [||G(F(y))-y||_1]
\end{equation}
The combined objective function is simply:
\begin{equation}
    \mathcal{L}(G,F,D_X,D_Y) = \mathcal{L}_{GAN}(G,D_Y,X,Y) + \mathcal{L}_{GAN}(F,D_X,Y,X) +  \lambda \cdot \mathcal{L}_{cyc}(G,F),
\end{equation}
where $\lambda$ is a tuning parameter to weight the relative importance of the networks \textbf{G} and \textbf{F}. Enforcing cycle consistency ensures that the data produced by the networks is statistically representative of and bounded by the training dataset, a property that makes the architecture suitable for use in seismic inversion.
\section{Convolutional Synthetic Seismic Data}
\begin{figure}[!htb]
    \centering
    \includegraphics[width=\textwidth]{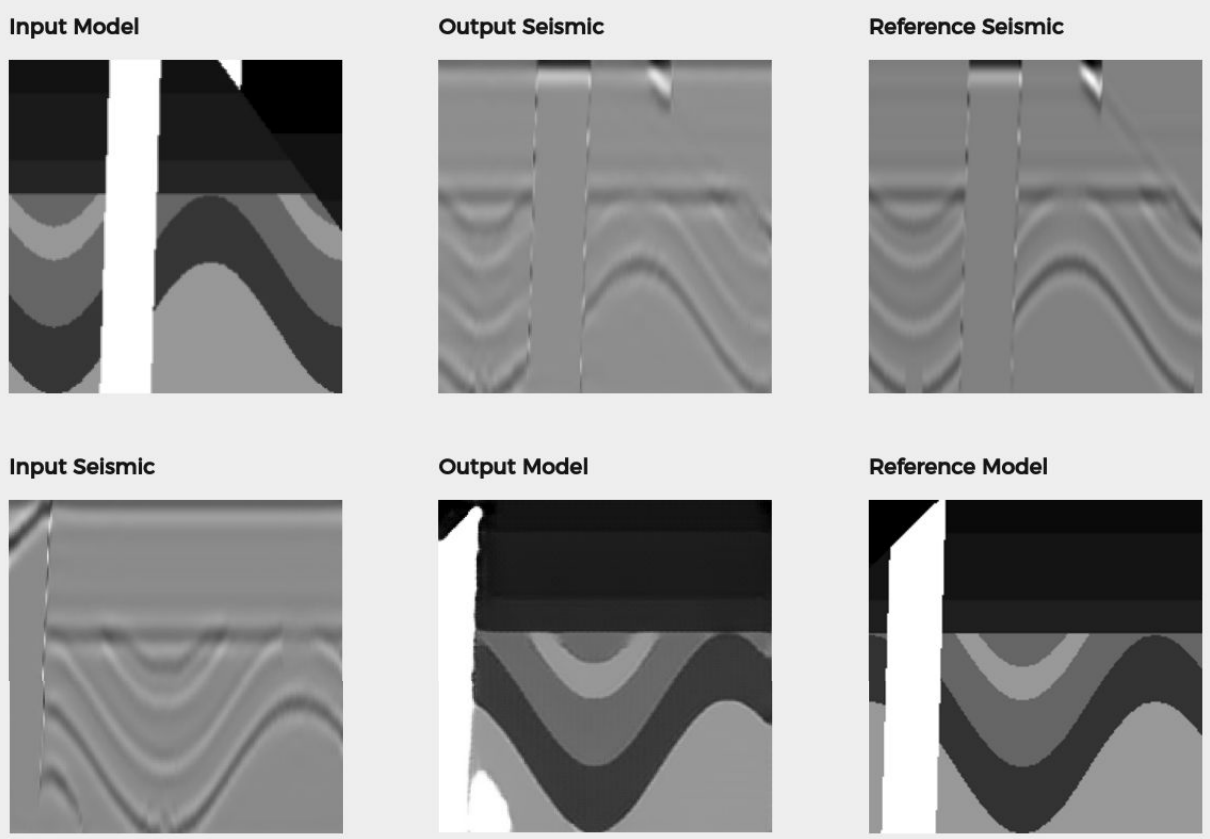}
    \caption{We evaluate the performance of the forward model \textbf{G} and inverse operator \textbf{F} based on unseen data of a dyke-anticline training dataset. The first row shows the forward pass through network \textbf{G} from the velocity model to the seismic domain. Row 2 shows the forward pass of network \textbf{F} from the seismic to the model domain. In both cases excellent agreement can be found with the reference images (Column 3).}
    \label{fig:convtrain}
\end{figure}
Initially, the network has been tested on geological models with a variety of features. We use a geological modelling package to generate realistic model data with multiple layers with varying velocity/impedance and thickness, folding, faulting and dyke intrusions. The synthetic seismic was generated by convolving the associated reflectivity with a Ricker wavelet.

Figure \ref{fig:convtrain} shows the input, result and reference for the generative networks \textbf{G} and \textbf{F}. The first row shows the forward pass from the model domain to seismic domain. A comparison with reference data shows a good match in both the structure and amplitudes.

\section{Marmousi2 Synthetic Seismic Data}
True seismic with its associated velocity errors and noise-related artifacts presents a much bigger challenge compared to the first example. To demonstrate the robustness of the method, we train on synthetic pre-stack Kirchhoff depth migrated seismic of the (elastic) Marmousi2 dataset \citep{martin2004marmousi2}. The data patches extracted from the data show much greater variability and less bias towards high velocity dykes that were prevalent in the convolutional synthetic seismic. 

\begin{figure}[]
    \centering
    \includegraphics[width=\textwidth]{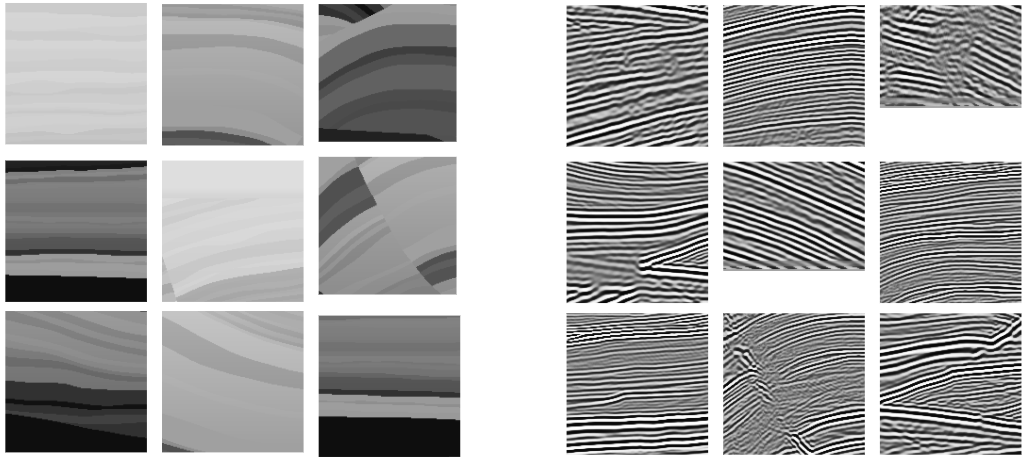}
    \caption{Example training patches extracted from the Marmousi2 synthetic model by \citet{martin2004marmousi2}. On the left velocities and on the right synthetic seismic forward models are shown.}
    \label{fig:marmousi2train}
\end{figure}

The training was further improved by two pre-processing steps: contrast-enhancement and histogram equalization. The 2D patches we extract from the model, processed and fed to the network are shown in figure \ref{fig:marmousi2train}. 

\section{Results}
\begin{figure}[h]
    \centering
    \includegraphics[width=\textwidth]{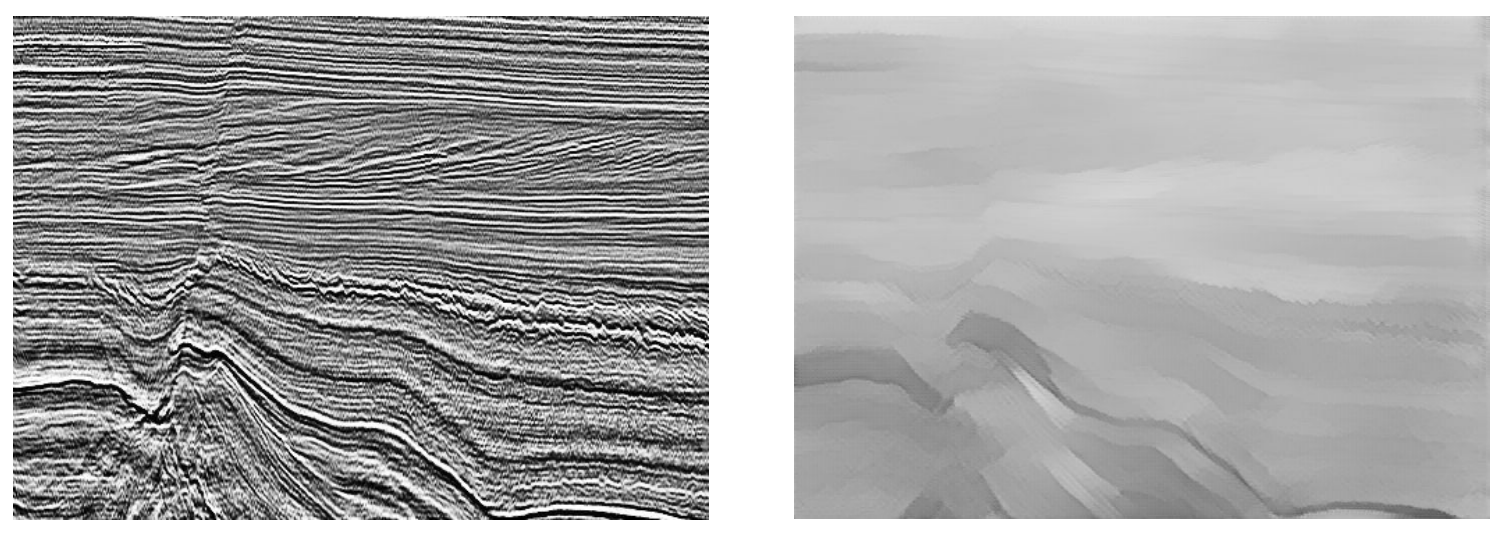}
    \caption{(left) Input seismic to network \textbf{F} to test the seismic inversion performance of generative adversarial networks. (right) Extracted velocity model generated by network \textbf{F}.}
    \label{fig:f3in}
\end{figure}

We test the improved network \textbf{F} by taking real data ("Dutch F3", left panel figure \ref{fig:f3in}) as input, that the network has never seen before. The section suffers from migration artifacts (bottom left), and occasional non-continuous reflectors, often a problem for computer vision algorithms. Low contrast regions on top of a high contrast region shows different internal structures and geometries that the network likely has not seen before.

Figure \ref{fig:f3in} shows the result of the mapping process of network \textbf{F}. The run-time of this seismic inversion process (the network's operation (\textbf{F})) is in the order of seconds (GPU time). High contrast areas from the seismic have been identified accordingly. The generated model shows large velocity contrasts where strong reflections occur and changes more smoothly otherwise. The fault is preserved in the velocity model, while the velocity model shows some continuity of velocities across the fault where appropriate.

\section{Conclusions}

We have presented a method to generalise seismic forward and inverse modeling approaches using domain transfer methods. Sets of training images of two-dimensional synthetic velocity models and forward models have been used to train a pair of deep convolutional neural networks. Once trained, these networks allow extremely fast extraction of estimated velocity fields and geological structure showing qualitatively good results on unseen seismic observations such as the F3 dataset. In our experimentation so far the technique appears to be particularly robust even when training is performed on synthetic datasets containing velocity errors and noise artifacts, providing convincing forward pass results on seismic data from the field. We believe the cyclic consistency constraint within the architecture and the associated relaxation of the requirement of perfectly matched paired input-output images plays a key role in stabilising the network, making this transfer possible.

\section{Acknowledgements}
The authors would like to thank the University of Houston for the Marmousi2 data and DGB for the dutch F3 data set. \citet{zhu2017unpaired} for the open source code release. Agile* Scientific and Total for organizing the 2017 subsurface hackathon that sparked this collaboration.

\bibliography{bib}

\end{document}